\documentstyle[preprint,aps]{revtex}
\tighten
\begin{document}

\title{ Eta-meson light nucleus scattering and Charge Symmetry
Breaking\footnote{
   Talk given at the {\it International Workshop on Exciting Physics with
New Accelerators Facilities (EXPAF97)}, SPring8,  Aioi,
Japan, March 11-13, 1997.
}}
\author{S. A. Sofianos, S. A. Rakityansky, and M. Braun}
\address{ Physics Department, University of South Africa,
 P.O.Box 392,Pretoria 0001, South Africa}
\maketitle
\vspace{0.8cm}

A considerable number of experiments on $\eta$--nucleon  and $\eta$--nucleus
interaction have been performed in recent years and  in parallel a number of
theoretical works which employed various models to handle the $\eta$-nucleus
dynamics have been published. Such a surge of interest is
attributed to  at least two main reasons. The first is of fundamental
character and addresses questions of particle physics such as the  breakdown
of the Okubo--Zweig--Iizuka (OZI) rule and  the Charge--Symmetry Breaking
(CSB) problem. The second reason comes from
nuclear physics. The idea of forming an $\eta$--mesic nucleus is very
attractive to both  experimentalists and theorists who are facing
the challenge to explore  new phenomena  and address questions related
to the great deal of overlapping between Nuclear and Particle Physics.

In general the $\eta$--physics is similar in spirit to a number of
ideas which have been vigorously pursued in the last couple of decades:
$\Lambda(\Sigma)$  hypernuclear states, dibaryon resonances,
antiprotonic states, just to mention a  few. However, the $\eta$--meson
has a very short lifetime ($\sim 10^{-19}$~sec) which makes it impossible to
generate $\eta$--beams. Due to it $\eta$--mesons are available only in
final states of certain nuclear reactions. And the $\eta$-nucleus
system plays an indispensable role in such investigations.

For example the OZI rule can, in principle, be tested by examining the
hidden strange--quark ($s\bar s$) component of the nucleon. For
this purpose, an experiment involving
simultaneous measurements of $\eta$ and $\eta'$ meson scattering from
a single nucleon has been proposed \cite{dover}.  However, the
feasibility of such experiment is limited by the lifetime  of the mesons
and the only possibility to circumvent this
difficulty is to try to extract information on $\eta N$ and
$\eta' N$ interactions using mesons produced on one nucleon
and scattered by other nucleons. This leads us  from the $\eta N$
problem to the  $\eta$--nucleus one.

The CSB question has been the subject of numerous
investigations in nuclear and particle physics.
One such example from nuclear physics is  a possible manifestation of
a CSB in the mirror photoproton and photoneutron
reactions $^4$He($\gamma$,p)$^3$H and $^4$He($\gamma$,n)$^3$He which
puzzled the experimentalists and the theoreticians alike, raising
controversies for several decades without the question being settled
yet (see \cite{photo} and Refs. therein).

It is nowadays believed that the CSB is caused by the mass difference
between the $u$ and $d$ quarks which, unfortunately, cannot be measured
directly. However, it can be formulated in another way,
namely, in terms of quantum mixing of $\eta$ and $\pi^0$ mesons which have
different isospins,
\begin{eqnarray}
\nonumber
|\eta^{\phantom{0}}>&=&|\tilde\pi>\sin\theta+|\tilde\eta>\cos\theta \ ,\\
     |\pi^0> &=&|\tilde \pi> \cos\theta -  |\tilde \eta>\sin\theta \ ,
\label{mix}
\end{eqnarray}
where $|\tilde \eta>$ and $|\tilde \pi>$ are pure isostopic states.
When this mixing is taken into account along with the electromagnetic
corrections one can reproduce the isomultiplet mass differences of ($n-p$),
($\Xi^--\Xi^0$), and ($\Sigma^--\Sigma^+$) pairs  with a mixing angle of
$\theta=0.013$ \cite{Mol}. This mixing can also manifest itself
in  reactions involving these particles. One such reaction has been
recently observed experimentally by Goldzahl et al. \cite{Gold}
in the fusion process
\begin{equation}
\label{ddapi}
	d+d\rightarrow \alpha+\pi^0
\end{equation}
at an incident deutron energy $T_d=1100$\,MeV where a cross--section
of 0.97\,pb/sr was found which is much higher than the one
expected from pure electromagnetic $\pi^0$ production in the $dd$
collision.
Since the $d$ and $\alpha$ have isospin zero and the $\pi^0$ 1,
it is obvious that this reaction is forbidden except if CSB is
manifested. Such a manifestation can be attributed to a mutual
transformation  of $\pi^0$ and $\eta$, i.e., the
reaction can proceed via the formation of an $\eta$-meson in an
intermediate step,
\begin{equation}
\label{ddeta}
\begin{array}{ccl}
	 & & \alpha+\eta \\
	d+d\rightarrow \alpha+\tilde\eta & {\nearrow\atop\searrow} & \\
	& & \alpha+\pi^0\,,
\end{array}
\end{equation}
where the zero--isospin particle $\tilde \eta$  is a mixture of the
physical $\eta$ and $\pi^0$  states according to Eq. (\ref{mix}),
$$
  |\tilde\eta> = \frac{1}{\sqrt{1+\lambda^2}}(|\eta> -\lambda |\pi^0>)\ ,
$$
with $\lambda=\tan\theta$. In such a case the reaction amplitudes
for (\ref{ddeta}) are mutually proportional
\begin{equation}
\label{flf}
	f(dd\rightarrow\alpha\pi^0)\approx -\lambda
	f(dd\rightarrow\alpha\eta)\ .
\end{equation}
Based on these ideas Coon and Preedom \cite{Coon} predicted earlier a
cross section for the reaction (\ref{ddapi}) of 0.12\,pb/sr,
i.e., about 10 time less than the experimental value.

To explain this discrepancy, Wilkin \cite{Wilkin} surmised that the
amplitude $f(dd\rightarrow\alpha\eta)$ is enhanced due to the resonant
character of the $\eta\alpha$ interaction in the final state. Then an
enhancement of the right hand side of Eq. (\ref{flf}) automatically causes
an increase of its left hand side. Following this idea, he incorporated the
final state interaction  via the rudimentary formula
\begin{equation}
\label{wil}
	f(dd\rightarrow\alpha\eta)=\frac{{\rm const}}
	{1-ia_{\eta\alpha}p} \ ,
\end{equation}
which involves the $\eta\alpha$ scattering length $a_{\eta\alpha}$ and the
relative momentum $p$, and adjusted the constant in Eq. (\ref{wil}) to fit
few experimental points available for the reaction
$dd\rightarrow\alpha\eta$ \cite{Fras} and thereby obtained the
cross--section for (\ref{ddapi}).

Such a simplified approach, however, can be considered only as a first
qualitative step towards a proper understanding of the phenomenon. A
quantitative analysis should be based on a microscopic treatment of the
few--body dynamics of the $\eta {}^4{\rm He}$ system which requires
the development of models appropriate
for the description of the $\eta$--meson interaction with
light nuclei.

The  $\eta d$ problem has been considered for the first time
 on the basis of the exact Faddeev
equations in \cite{edfadd}. The use of this rigorous method, however,
is limited to four particles.   An alternative microscopic approach
is the Finite Rank Approximation (FRA) of the
nuclear Hamiltonian, within which the $\eta d$, $\eta {}^3{\rm H}$,
$\eta {}^3{\rm He}$, and $\eta {}^4{\rm He}$ systems were considered
(see Refs. \cite{Raki1,Raki2,Bel1,Raki3,Raki4,Raki5}). The only
approximation used in this approach is the truncation of the spectral
expansion of the nuclear
Hamiltonian to the effect that only the ground state of the nucleus
is retained,
\begin{equation}
\label{ha}
	H_A \approx {\cal E}_0|\psi_0\rangle\langle\psi_0| \ .
\end{equation}
Such an approximation, known in the scattering theory as the coherent
approximation, means that we neglect virtual
excitations of the nuclear target during the $\eta$--meson scattering.
Of course when the collision energy is close to or above the excitation
threshold, the accuracy of such an approach is questionable. If, however,
we are interested in  very low energies (scattering length calculations)
or negative energies (bound states), then this approximation
(\ref{ha}) must be very reliable. Although a formal proof of this was
given in Ref.\cite{pup},  from our physical intuition it is clear
that  contributions to $\eta {}^4{\rm He}$ scattering length, for example,
from virtual excitations which lie at
least 20\,MeV above threshold should be very small.

In the FRA the  approximation (\ref{ha}) is the only one used while
the rest of the calculations can be done exactly. To derive the
FRA equations, we
start from the many--body Lippmann--Schwinger equation for the total
$T$--operator, and after certain algebraic manipulations we obtain the
following equivalent system of equations
\begin{equation}
\label{opeq}
\begin{array}{ccl}
T(z)&=&\sum\limits_{i=1}^AT_i^0(z)+\sum\limits_{i=1}^AT_i^0(z)
{\displaystyle\frac{1}{z-H_0}}
     H_A{\displaystyle\frac{1}{z-H_0-H_A}}T(z) \ ,\\
&&\\
T_i^0(z)&=&t_i(z)+t_i(z){\displaystyle\frac{1}{z-H_0}}
\sum\limits_{j\ne i}^AT_j^0(z) \ ,
\end{array}
\end{equation}
where $A$ is the number of nucleons, $z$ the total $\eta A$--energy, $H_0$
the Hamiltonian of the free motion of $\eta$ with respect to the $c.m.$ of the
nucleus, $T_i^0$ are the auxiliary operators, and $t_i$ the two--body
$\eta$-nucleon $T$--operators in the many--body space. With the
approximation (\ref{ha}) the integral equation
for the elastic scattering amplitude
$\langle{\vec {p'}},\psi_0|T(z)|{\vec p},\psi_0\rangle$
becomes one--dimensional (only with respect to the relative momentum $p$)
\begin{equation}
\label{avereq}
\langle\psi_0|T(z)|\psi_0\rangle=
\langle\psi_0|\sum\limits_{i=1}^AT_i^0(z)|\psi_0\rangle+
\langle\psi_0|\sum\limits_{i=1}^AT_i^0(z)|\psi_0\rangle
\frac{{\cal E}_0}{(z-H_0)(z-H_0-{\cal E}_0)}\langle\psi_0|
T(z)|\psi_0\rangle \ .
\end{equation}
Thus, the whole problem is separated into two parts: first we have to find the
auxiliary operators $T_i^0$ (in the many--body space) and then to solve the
integral equation (\ref{avereq}) (in the $p$--space only). This separation
has been chosen in such way that the equations in the many--body space (for
$T_i^0$) do not involve $H_A$ and can therefore be solved exactly for
reasonably large number of nucleons  $A$, especially with a
separable two--body $\eta N$ interaction. The nuclear wave function
$\psi_0$ can be found separately by any appropriate method.

Since direct experiments on $\eta N$ scattering are not feasible, our
knowledge of the  $\eta N$ interaction is poor.
At low energies, however, it has been established  that
 this interaction is attractive and very strong
due to the dominance of the $N^*(1535)$   $S_{11}$ - resonance.
This dominance implies
that the $\eta N$ amplitude has a pole at a complex energy and can be
represented in the  form
\begin{equation}
\label{ten}
	   t_{\eta N}(p',p;z) = \frac {\lambda}{(p'^2+
	   \alpha^2)(z - E_0 + i\Gamma/2)(p^2+\alpha^2)}
\end{equation}
with $E_0 = 1535\; {\rm MeV} - ( m_N + m_{\eta} )$, $\Gamma = 150\;
{\rm MeV}\;$ \cite{PDGr}. It was found via a two--channel fit to the
$\pi N\rightarrow\pi N$ and $\pi N\rightarrow\eta N$ experimental data that
the  range parameter is $\alpha =2.357$~fm$^{-1}$ \cite{bhal}. The remaining
parameter $\lambda$ is chosen   to provide the correct zero-energy on-shell
limit, i.e., to reproduce the $\eta N$ scattering length $a_{\eta N}$,
$$
 t_{\eta  N}(0,0,0) = - \frac {2\pi}{\mu_{\eta  N}}a_{\eta  N}.
$$
The scattering length $a_{\eta N}$ is not  accurately known.
Different analyses \cite{batinic} provided values for the real part in the
range
${\rm Re\,}a_{\eta N}\in [0.27,0.98]\, {\rm fm}$ and for the imaginary part
${\rm Im\,}a_{\eta N}\in [0.19,0.37]\,{\rm fm}$ .

Using the Marchenko inverse scattering theory, we have constructed a local
$\eta N$ potential which generates the same phase shifts as the amplitude
(\ref{ten}) with $a_{\eta N}=(0.55+i0.30)$ fm. This potential is
depicted in Fig. 1 where the $\eta N$ attraction is clearly seen.
Since there is no centrifugal barrier in the $S$--wave state, the repulsive
barrier of the potential itself is responsible for the $S_{11}$  resonance
at $\sim 50$ MeV above threshold in the attractive well. This barrier is not
wide and therefore easily penetrable causing the large width of the
resonance.

The strong $\eta N$ attraction raises the question of a possible existence of
$\eta A$ bound states. If they do exist, the final--state
$\eta$--meson can be trapped by the nucleus for a relatively long time. This
would open up new avenues for the elucidation of the $\eta A$
dynamics at low energies and perhaps shed more light on the
CSB  problem. First estimates obtained in the framework of the
optical potential theory
put a lower bound on the atomic number $A$ for which a
bound state could exist, namely, $A\ge12$ \cite{Haider}.
Strictly speaking, genuine bound states cannot exist in $\eta$--nucleus
systems because the inelastic channel $\eta N\rightarrow N\pi$ is always
open. This makes the $\eta N$ Hamiltonian non-Hermitian and thus
only  quasi--bound states can exist.

In order to obtain reliable scattering lengths and to shed light on the
quasi--bound state problem, we have performed  microscopic
 calculations of the $\eta$ scattering from ${}^2{\rm H}$,
${}^3{\rm H}$, ${}^3{\rm He}$, and ${}^4{\rm He}$ nuclei in the
framework of the FRA (see Refs. \cite{Raki1,Raki2,Bel1,Raki3,Raki4,Raki5}).
To obtain the necessary nuclear wave
functions $\psi_0$ we employed the integro--differential equations
(IDEA) \cite{IDEA} which, with the $S$--wave Malfliet--Tjon $NN$--potential
used, coincide with the exact Faddeev equations. The scattering lengths thus
obtained are shown on Fig. 2 ($a_{\eta d}$ and $a_{\eta {}^4{\rm He}}$)
and Fig. 3 ($a_{\eta {}^3{\rm H}}$ and $a_{\eta {}^3{\rm He}}$).

For each of the four nuclei considered, the scattering lengths
were calculated with eight values of the strength parameter
$\lambda$ of Eq. (\ref{ten}),
corresponding to eight values of ${\rm Re\,}a_{\eta N}$:
$\{(0.2+0.1n)\ {\rm fm}; n=1,8\}$, which cover the uncertainty
interval. The ${\rm Im\,}a_{\eta N}$ was fixed to the value 0.3 fm.
An increase of ${\rm Re\,}a_{\eta N}$ moves the points on Figs. 2 and 3
along the curve trajectories in the anti-clockwise direction. When
${\rm Re\,}a_{\eta N}$ exceeds a certain critical value the $\eta N$
interaction becomes strong enough to generate a quasi--bound state.
The points on the corresponding trajectories beyond that value are
shown by filled circles (the trajectories for $^3$He and $^3$H
are practically the same).

\begin{center}
\unitlength=0.5mm
\begin{picture}(280,110)
\put(25,80){%
\begin{picture}(0,0)%
\put(0,0){\line(1,0){50}}
\put(0,-60){\line(0,1){80}}
\multiput(5,-2)(5,0){10}{\line(0,1){2}}
\put(25,-4){\line(0,1){2}}
\put(50,-4){\line(0,1){2}}
\multiput(-2,-60)(0,10){9}{\line(1,0){2}}
\put(-4,-50){\line(1,0){2}}
\put(50,-22){\llap{$r$ (fm)}}
\put(-4,15){\llap{$V_{\eta N}$ (MeV)}}
\put(15,-50){Im}
\put(15,-48){\vector(-2,1){10}}
\put(30,15){Re}
\put(30,17){\vector(-2,-1){10}}
\put(-5,-2){\llap{0}}
\put(-5,-52){\llap{-500}}
\put(48,-11){1}
\put(20,-11){0.5}
\put(-10,-80){Fig. 1}
\put(  .12500,-32.952){\circle*{0.1}}\put(  .12500,-56.017){\circle*{0.1}}
\put(  .25000,-32.974){\circle*{0.1}}\put(  .25000,-56.044){\circle*{0.1}}
\put(  .37500,-33.018){\circle*{0.1}}\put(  .37500,-55.988){\circle*{0.1}}
\put(  .50000,-33.080){\circle*{0.1}}\put(  .50000,-55.852){\circle*{0.1}}
\put(  .62500,-33.158){\circle*{0.1}}\put(  .62500,-55.637){\circle*{0.1}}
\put(  .75000,-33.247){\circle*{0.1}}\put(  .75000,-55.346){\circle*{0.1}}
\put(  .87500,-33.343){\circle*{0.1}}\put(  .87500,-54.981){\circle*{0.1}}
\put( 1.00000,-33.445){\circle*{0.1}}\put( 1.00000,-54.545){\circle*{0.1}}
\put( 1.12500,-33.549){\circle*{0.1}}\put( 1.12500,-54.042){\circle*{0.1}}
\put( 1.25000,-33.652){\circle*{0.1}}\put( 1.25000,-53.472){\circle*{0.1}}
\put( 1.37500,-33.753){\circle*{0.1}}\put( 1.37500,-52.841){\circle*{0.1}}
\put( 1.50000,-33.847){\circle*{0.1}}\put( 1.50000,-52.150){\circle*{0.1}}
\put( 1.62500,-33.934){\circle*{0.1}}\put( 1.62500,-51.403){\circle*{0.1}}
\put( 1.75000,-34.011){\circle*{0.1}}\put( 1.75000,-50.603){\circle*{0.1}}
\put( 1.87500,-34.076){\circle*{0.1}}\put( 1.87500,-49.753){\circle*{0.1}}
\put( 2.00000,-34.127){\circle*{0.1}}\put( 2.00000,-48.857){\circle*{0.1}}
\put( 2.12500,-34.161){\circle*{0.1}}\put( 2.12500,-47.917){\circle*{0.1}}
\put( 2.25000,-34.178){\circle*{0.1}}\put( 2.25000,-46.936){\circle*{0.1}}
\put( 2.37500,-34.176){\circle*{0.1}}\put( 2.37500,-45.918){\circle*{0.1}}
\put( 2.50000,-34.153){\circle*{0.1}}\put( 2.50000,-44.865){\circle*{0.1}}
\put( 2.62500,-34.107){\circle*{0.1}}\put( 2.62500,-43.781){\circle*{0.1}}
\put( 2.75000,-34.039){\circle*{0.1}}\put( 2.75000,-42.668){\circle*{0.1}}
\put( 2.87500,-33.946){\circle*{0.1}}\put( 2.87500,-41.529){\circle*{0.1}}
\put( 3.00000,-33.829){\circle*{0.1}}\put( 3.00000,-40.367){\circle*{0.1}}
\put( 3.12500,-33.686){\circle*{0.1}}\put( 3.12500,-39.185){\circle*{0.1}}
\put( 3.25000,-33.517){\circle*{0.1}}\put( 3.25000,-37.985){\circle*{0.1}}
\put( 3.37500,-33.321){\circle*{0.1}}\put( 3.37500,-36.771){\circle*{0.1}}
\put( 3.50000,-33.099){\circle*{0.1}}\put( 3.50000,-35.544){\circle*{0.1}}
\put( 3.62500,-32.850){\circle*{0.1}}\put( 3.62500,-34.307){\circle*{0.1}}
\put( 3.75000,-32.574){\circle*{0.1}}\put( 3.75000,-33.064){\circle*{0.1}}
\put( 3.87500,-32.271){\circle*{0.1}}\put( 3.87500,-31.816){\circle*{0.1}}
\put( 4.00000,-31.942){\circle*{0.1}}\put( 4.00000,-30.566){\circle*{0.1}}
\put( 4.12500,-31.586){\circle*{0.1}}\put( 4.12500,-29.316){\circle*{0.1}}
\put( 4.25000,-31.204){\circle*{0.1}}\put( 4.25000,-28.068){\circle*{0.1}}
\put( 4.37500,-30.797){\circle*{0.1}}\put( 4.37500,-26.824){\circle*{0.1}}
\put( 4.50000,-30.364){\circle*{0.1}}\put( 4.50000,-25.588){\circle*{0.1}}
\put( 4.62500,-29.907){\circle*{0.1}}\put( 4.62500,-24.359){\circle*{0.1}}
\put( 4.75000,-29.426){\circle*{0.1}}\put( 4.75000,-23.142){\circle*{0.1}}
\put( 4.87500,-28.921){\circle*{0.1}}\put( 4.87500,-21.936){\circle*{0.1}}
\put( 5.00000,-28.394){\circle*{0.1}}\put( 5.00000,-20.744){\circle*{0.1}}
\put( 5.12500,-27.846){\circle*{0.1}}\put( 5.12500,-19.568){\circle*{0.1}}
\put( 5.25000,-27.277){\circle*{0.1}}\put( 5.25000,-18.408){\circle*{0.1}}
\put( 5.37500,-26.688){\circle*{0.1}}\put( 5.37500,-17.267){\circle*{0.1}}
\put( 5.50000,-26.081){\circle*{0.1}}\put( 5.50000,-16.145){\circle*{0.1}}
\put( 5.62500,-25.455){\circle*{0.1}}\put( 5.62500,-15.045){\circle*{0.1}}
\put( 5.75000,-24.814){\circle*{0.1}}\put( 5.75000,-13.966){\circle*{0.1}}
\put( 5.87500,-24.156){\circle*{0.1}}\put( 5.87500,-12.910){\circle*{0.1}}
\put( 6.00000,-23.484){\circle*{0.1}}\put( 6.00000,-11.879){\circle*{0.1}}
\put( 6.12500,-22.799){\circle*{0.1}}\put( 6.12500,-10.872){\circle*{0.1}}
\put( 6.25000,-22.102){\circle*{0.1}}\put( 6.25000, -9.890){\circle*{0.1}}
\put( 6.37500,-21.394){\circle*{0.1}}\put( 6.37500, -8.935){\circle*{0.1}}
\put( 6.50000,-20.676){\circle*{0.1}}\put( 6.50000, -8.007){\circle*{0.1}}
\put( 6.62500,-19.950){\circle*{0.1}}\put( 6.62500, -7.107){\circle*{0.1}}
\put( 6.75000,-19.216){\circle*{0.1}}\put( 6.75000, -6.234){\circle*{0.1}}
\put( 6.87500,-18.476){\circle*{0.1}}\put( 6.87500, -5.390){\circle*{0.1}}
\put( 7.00000,-17.730){\circle*{0.1}}\put( 7.00000, -4.574){\circle*{0.1}}
\put( 7.12500,-16.981){\circle*{0.1}}\put( 7.12500, -3.787){\circle*{0.1}}
\put( 7.25000,-16.228){\circle*{0.1}}\put( 7.25000, -3.028){\circle*{0.1}}
\put( 7.37500,-15.474){\circle*{0.1}}\put( 7.37500, -2.299){\circle*{0.1}}
\put( 7.50000,-14.719){\circle*{0.1}}\put( 7.50000, -1.599){\circle*{0.1}}
\put( 7.62500,-13.965){\circle*{0.1}}\put( 7.62500,  -.927){\circle*{0.1}}
\put( 7.75000,-13.211){\circle*{0.1}}\put( 7.75000,  -.285){\circle*{0.1}}
\put( 7.87500,-12.460){\circle*{0.1}}\put( 7.87500,   .330){\circle*{0.1}}
\put( 8.00000,-11.713){\circle*{0.1}}\put( 8.00000,   .915){\circle*{0.1}}
\put( 8.12500,-10.969){\circle*{0.1}}\put( 8.12500,  1.473){\circle*{0.1}}
\put( 8.25000,-10.231){\circle*{0.1}}\put( 8.25000,  2.003){\circle*{0.1}}
\put( 8.37500, -9.499){\circle*{0.1}}\put( 8.37500,  2.505){\circle*{0.1}}
\put( 8.50000, -8.774){\circle*{0.1}}\put( 8.50000,  2.980){\circle*{0.1}}
\put( 8.62500, -8.057){\circle*{0.1}}\put( 8.62500,  3.429){\circle*{0.1}}
\put( 8.75000, -7.348){\circle*{0.1}}\put( 8.75000,  3.852){\circle*{0.1}}
\put( 8.87500, -6.648){\circle*{0.1}}\put( 8.87500,  4.249){\circle*{0.1}}
\put( 9.00000, -5.958){\circle*{0.1}}\put( 9.00000,  4.621){\circle*{0.1}}
\put( 9.12500, -5.279){\circle*{0.1}}\put( 9.12500,  4.969){\circle*{0.1}}
\put( 9.25000, -4.611){\circle*{0.1}}\put( 9.25000,  5.293){\circle*{0.1}}
\put( 9.37500, -3.954){\circle*{0.1}}\put( 9.37500,  5.594){\circle*{0.1}}
\put( 9.50000, -3.310){\circle*{0.1}}\put( 9.50000,  5.872){\circle*{0.1}}
\put( 9.62500, -2.679){\circle*{0.1}}\put( 9.62500,  6.129){\circle*{0.1}}
\put( 9.75000, -2.061){\circle*{0.1}}\put( 9.75000,  6.364){\circle*{0.1}}
\put( 9.87500, -1.456){\circle*{0.1}}\put( 9.87500,  6.579){\circle*{0.1}}
\put(10.00000,  -.866){\circle*{0.1}}\put(10.00000,  6.774){\circle*{0.1}}
\put(10.12500,  -.290){\circle*{0.1}}\put(10.12500,  6.949){\circle*{0.1}}
\put(10.25000,   .272){\circle*{0.1}}\put(10.25000,  7.107){\circle*{0.1}}
\put(10.37500,   .818){\circle*{0.1}}\put(10.37500,  7.247){\circle*{0.1}}
\put(10.50000,  1.349){\circle*{0.1}}\put(10.50000,  7.370){\circle*{0.1}}
\put(10.62500,  1.865){\circle*{0.1}}\put(10.62500,  7.476){\circle*{0.1}}
\put(10.75000,  2.366){\circle*{0.1}}\put(10.75000,  7.568){\circle*{0.1}}
\put(10.87500,  2.850){\circle*{0.1}}\put(10.87500,  7.644){\circle*{0.1}}
\put(11.00000,  3.319){\circle*{0.1}}\put(11.00000,  7.707){\circle*{0.1}}
\put(11.12500,  3.772){\circle*{0.1}}\put(11.12500,  7.756){\circle*{0.1}}
\put(11.25000,  4.209){\circle*{0.1}}\put(11.25000,  7.792){\circle*{0.1}}
\put(11.37500,  4.630){\circle*{0.1}}\put(11.37500,  7.817){\circle*{0.1}}
\put(11.50000,  5.036){\circle*{0.1}}\put(11.50000,  7.830){\circle*{0.1}}
\put(11.62500,  5.425){\circle*{0.1}}\put(11.62500,  7.833){\circle*{0.1}}
\put(11.75000,  5.799){\circle*{0.1}}\put(11.75000,  7.826){\circle*{0.1}}
\put(11.87500,  6.157){\circle*{0.1}}\put(11.87500,  7.810){\circle*{0.1}}
\put(12.00000,  6.499){\circle*{0.1}}\put(12.00000,  7.785){\circle*{0.1}}
\put(12.12500,  6.826){\circle*{0.1}}\put(12.12500,  7.752){\circle*{0.1}}
\put(12.25000,  7.138){\circle*{0.1}}\put(12.25000,  7.711){\circle*{0.1}}
\put(12.37500,  7.434){\circle*{0.1}}\put(12.37500,  7.664){\circle*{0.1}}
\put(12.50000,  7.716){\circle*{0.1}}\put(12.50000,  7.610){\circle*{0.1}}
\put(12.62500,  7.983){\circle*{0.1}}\put(12.62500,  7.551){\circle*{0.1}}
\put(12.75000,  8.236){\circle*{0.1}}\put(12.75000,  7.487){\circle*{0.1}}
\put(12.87500,  8.474){\circle*{0.1}}\put(12.87500,  7.418){\circle*{0.1}}
\put(13.00000,  8.698){\circle*{0.1}}\put(13.00000,  7.345){\circle*{0.1}}
\put(13.12500,  8.909){\circle*{0.1}}\put(13.12500,  7.268){\circle*{0.1}}
\put(13.25000,  9.106){\circle*{0.1}}\put(13.25000,  7.188){\circle*{0.1}}
\put(13.37500,  9.290){\circle*{0.1}}\put(13.37500,  7.106){\circle*{0.1}}
\put(13.50000,  9.462){\circle*{0.1}}\put(13.50000,  7.021){\circle*{0.1}}
\put(13.62500,  9.621){\circle*{0.1}}\put(13.62500,  6.934){\circle*{0.1}}
\put(13.75000,  9.767){\circle*{0.1}}\put(13.75000,  6.846){\circle*{0.1}}
\put(13.87500,  9.902){\circle*{0.1}}\put(13.87500,  6.757){\circle*{0.1}}
\put(14.00000, 10.025){\circle*{0.1}}\put(14.00000,  6.667){\circle*{0.1}}
\put(14.12500, 10.138){\circle*{0.1}}\put(14.12500,  6.576){\circle*{0.1}}
\put(14.25000, 10.239){\circle*{0.1}}\put(14.25000,  6.486){\circle*{0.1}}
\put(14.37500, 10.330){\circle*{0.1}}\put(14.37500,  6.396){\circle*{0.1}}
\put(14.50000, 10.410){\circle*{0.1}}\put(14.50000,  6.306){\circle*{0.1}}
\put(14.62500, 10.481){\circle*{0.1}}\put(14.62500,  6.217){\circle*{0.1}}
\put(14.75000, 10.543){\circle*{0.1}}\put(14.75000,  6.129){\circle*{0.1}}
\put(14.87500, 10.595){\circle*{0.1}}\put(14.87500,  6.042){\circle*{0.1}}
\put(15.00000, 10.638){\circle*{0.1}}\put(15.00000,  5.956){\circle*{0.1}}
\put(15.12500, 10.674){\circle*{0.1}}\put(15.12500,  5.873){\circle*{0.1}}
\put(15.25000, 10.701){\circle*{0.1}}\put(15.25000,  5.791){\circle*{0.1}}
\put(15.37500, 10.720){\circle*{0.1}}\put(15.37500,  5.711){\circle*{0.1}}
\put(15.50000, 10.732){\circle*{0.1}}\put(15.50000,  5.633){\circle*{0.1}}
\put(15.62500, 10.737){\circle*{0.1}}\put(15.62500,  5.558){\circle*{0.1}}
\put(15.75000, 10.735){\circle*{0.1}}\put(15.75000,  5.485){\circle*{0.1}}
\put(15.87500, 10.727){\circle*{0.1}}\put(15.87500,  5.415){\circle*{0.1}}
\put(16.00000, 10.713){\circle*{0.1}}\put(16.00000,  5.347){\circle*{0.1}}
\put(16.12500, 10.694){\circle*{0.1}}\put(16.12500,  5.282){\circle*{0.1}}
\put(16.25000, 10.668){\circle*{0.1}}\put(16.25000,  5.220){\circle*{0.1}}
\put(16.37500, 10.638){\circle*{0.1}}\put(16.37500,  5.161){\circle*{0.1}}
\put(16.50000, 10.603){\circle*{0.1}}\put(16.50000,  5.104){\circle*{0.1}}
\put(16.62500, 10.564){\circle*{0.1}}\put(16.62500,  5.051){\circle*{0.1}}
\put(16.75000, 10.520){\circle*{0.1}}\put(16.75000,  5.000){\circle*{0.1}}
\put(16.87500, 10.472){\circle*{0.1}}\put(16.87500,  4.953){\circle*{0.1}}
\put(17.00000, 10.421){\circle*{0.1}}\put(17.00000,  4.908){\circle*{0.1}}
\put(17.12500, 10.366){\circle*{0.1}}\put(17.12500,  4.867){\circle*{0.1}}
\put(17.25000, 10.309){\circle*{0.1}}\put(17.25000,  4.829){\circle*{0.1}}
\put(17.37500, 10.248){\circle*{0.1}}\put(17.37500,  4.793){\circle*{0.1}}
\put(17.50000, 10.185){\circle*{0.1}}\put(17.50000,  4.760){\circle*{0.1}}
\put(17.62500, 10.119){\circle*{0.1}}\put(17.62500,  4.731){\circle*{0.1}}
\put(17.75000, 10.052){\circle*{0.1}}\put(17.75000,  4.704){\circle*{0.1}}
\put(17.87500,  9.982){\circle*{0.1}}\put(17.87500,  4.680){\circle*{0.1}}
\put(18.00000,  9.911){\circle*{0.1}}\put(18.00000,  4.659){\circle*{0.1}}
\put(18.12500,  9.838){\circle*{0.1}}\put(18.12500,  4.640){\circle*{0.1}}
\put(18.25000,  9.764){\circle*{0.1}}\put(18.25000,  4.625){\circle*{0.1}}
\put(18.37500,  9.688){\circle*{0.1}}\put(18.37500,  4.611){\circle*{0.1}}
\put(18.50000,  9.612){\circle*{0.1}}\put(18.50000,  4.601){\circle*{0.1}}
\put(18.62500,  9.535){\circle*{0.1}}\put(18.62500,  4.593){\circle*{0.1}}
\put(18.75000,  9.458){\circle*{0.1}}\put(18.75000,  4.587){\circle*{0.1}}
\put(18.87500,  9.380){\circle*{0.1}}\put(18.87500,  4.583){\circle*{0.1}}
\put(19.00000,  9.302){\circle*{0.1}}\put(19.00000,  4.582){\circle*{0.1}}
\put(19.12500,  9.223){\circle*{0.1}}\put(19.12500,  4.582){\circle*{0.1}}
\put(19.25000,  9.145){\circle*{0.1}}\put(19.25000,  4.585){\circle*{0.1}}
\put(19.37500,  9.066){\circle*{0.1}}\put(19.37500,  4.589){\circle*{0.1}}
\put(19.50000,  8.988){\circle*{0.1}}\put(19.50000,  4.596){\circle*{0.1}}
\put(19.62500,  8.910){\circle*{0.1}}\put(19.62500,  4.604){\circle*{0.1}}
\put(19.75000,  8.833){\circle*{0.1}}\put(19.75000,  4.614){\circle*{0.1}}
\put(19.87500,  8.756){\circle*{0.1}}\put(19.87500,  4.625){\circle*{0.1}}
\put(20.00000,  8.680){\circle*{0.1}}\put(20.00000,  4.638){\circle*{0.1}}
\put(20.12500,  8.605){\circle*{0.1}}\put(20.12500,  4.652){\circle*{0.1}}
\put(20.25000,  8.530){\circle*{0.1}}\put(20.25000,  4.667){\circle*{0.1}}
\put(20.37500,  8.456){\circle*{0.1}}\put(20.37500,  4.684){\circle*{0.1}}
\put(20.50000,  8.383){\circle*{0.1}}\put(20.50000,  4.701){\circle*{0.1}}
\put(20.62500,  8.311){\circle*{0.1}}\put(20.62500,  4.720){\circle*{0.1}}
\put(20.75000,  8.240){\circle*{0.1}}\put(20.75000,  4.739){\circle*{0.1}}
\put(20.87500,  8.170){\circle*{0.1}}\put(20.87500,  4.759){\circle*{0.1}}
\put(21.00000,  8.101){\circle*{0.1}}\put(21.00000,  4.780){\circle*{0.1}}
\put(21.12500,  8.034){\circle*{0.1}}\put(21.12500,  4.802){\circle*{0.1}}
\put(21.25000,  7.967){\circle*{0.1}}\put(21.25000,  4.824){\circle*{0.1}}
\put(21.37500,  7.902){\circle*{0.1}}\put(21.37500,  4.846){\circle*{0.1}}
\put(21.50000,  7.838){\circle*{0.1}}\put(21.50000,  4.869){\circle*{0.1}}
\put(21.62500,  7.775){\circle*{0.1}}\put(21.62500,  4.892){\circle*{0.1}}
\put(21.75000,  7.713){\circle*{0.1}}\put(21.75000,  4.915){\circle*{0.1}}
\put(21.87500,  7.653){\circle*{0.1}}\put(21.87500,  4.938){\circle*{0.1}}
\put(22.00000,  7.594){\circle*{0.1}}\put(22.00000,  4.962){\circle*{0.1}}
\put(22.12500,  7.536){\circle*{0.1}}\put(22.12500,  4.985){\circle*{0.1}}
\put(22.25000,  7.479){\circle*{0.1}}\put(22.25000,  5.008){\circle*{0.1}}
\put(22.37500,  7.424){\circle*{0.1}}\put(22.37500,  5.031){\circle*{0.1}}
\put(22.50000,  7.370){\circle*{0.1}}\put(22.50000,  5.054){\circle*{0.1}}
\put(22.62500,  7.317){\circle*{0.1}}\put(22.62500,  5.076){\circle*{0.1}}
\put(22.75000,  7.265){\circle*{0.1}}\put(22.75000,  5.098){\circle*{0.1}}
\put(22.87500,  7.215){\circle*{0.1}}\put(22.87500,  5.120){\circle*{0.1}}
\put(23.00000,  7.165){\circle*{0.1}}\put(23.00000,  5.141){\circle*{0.1}}
\put(23.12500,  7.117){\circle*{0.1}}\put(23.12500,  5.162){\circle*{0.1}}
\put(23.25000,  7.070){\circle*{0.1}}\put(23.25000,  5.182){\circle*{0.1}}
\put(23.37500,  7.024){\circle*{0.1}}\put(23.37500,  5.201){\circle*{0.1}}
\put(23.50000,  6.979){\circle*{0.1}}\put(23.50000,  5.220){\circle*{0.1}}
\put(23.62500,  6.935){\circle*{0.1}}\put(23.62500,  5.238){\circle*{0.1}}
\put(23.75000,  6.892){\circle*{0.1}}\put(23.75000,  5.256){\circle*{0.1}}
\put(23.87500,  6.850){\circle*{0.1}}\put(23.87500,  5.272){\circle*{0.1}}
\put(24.00000,  6.809){\circle*{0.1}}\put(24.00000,  5.288){\circle*{0.1}}
\put(24.12500,  6.769){\circle*{0.1}}\put(24.12500,  5.303){\circle*{0.1}}
\put(24.25000,  6.729){\circle*{0.1}}\put(24.25000,  5.318){\circle*{0.1}}
\put(24.37500,  6.691){\circle*{0.1}}\put(24.37500,  5.331){\circle*{0.1}}
\put(24.50000,  6.653){\circle*{0.1}}\put(24.50000,  5.343){\circle*{0.1}}
\put(24.62500,  6.616){\circle*{0.1}}\put(24.62500,  5.355){\circle*{0.1}}
\put(24.75000,  6.580){\circle*{0.1}}\put(24.75000,  5.366){\circle*{0.1}}
\put(24.87500,  6.545){\circle*{0.1}}\put(24.87500,  5.375){\circle*{0.1}}
\put(25.00000,  6.510){\circle*{0.1}}\put(25.00000,  5.384){\circle*{0.1}}
\put(25.12500,  6.476){\circle*{0.1}}\put(25.12500,  5.392){\circle*{0.1}}
\put(25.25000,  6.442){\circle*{0.1}}\put(25.25000,  5.399){\circle*{0.1}}
\put(25.37500,  6.409){\circle*{0.1}}\put(25.37500,  5.405){\circle*{0.1}}
\put(25.50000,  6.376){\circle*{0.1}}\put(25.50000,  5.410){\circle*{0.1}}
\put(25.62500,  6.344){\circle*{0.1}}\put(25.62500,  5.414){\circle*{0.1}}
\put(25.75000,  6.312){\circle*{0.1}}\put(25.75000,  5.417){\circle*{0.1}}
\put(25.87500,  6.281){\circle*{0.1}}\put(25.87500,  5.419){\circle*{0.1}}
\put(26.00000,  6.250){\circle*{0.1}}\put(26.00000,  5.420){\circle*{0.1}}
\put(26.12500,  6.220){\circle*{0.1}}\put(26.12500,  5.420){\circle*{0.1}}
\put(26.25000,  6.189){\circle*{0.1}}\put(26.25000,  5.419){\circle*{0.1}}
\put(26.37500,  6.159){\circle*{0.1}}\put(26.37500,  5.417){\circle*{0.1}}
\put(26.50000,  6.130){\circle*{0.1}}\put(26.50000,  5.415){\circle*{0.1}}
\put(26.62500,  6.100){\circle*{0.1}}\put(26.62500,  5.411){\circle*{0.1}}
\put(26.75000,  6.071){\circle*{0.1}}\put(26.75000,  5.407){\circle*{0.1}}
\put(26.87500,  6.042){\circle*{0.1}}\put(26.87500,  5.401){\circle*{0.1}}
\put(27.00000,  6.013){\circle*{0.1}}\put(27.00000,  5.395){\circle*{0.1}}
\put(27.12500,  5.984){\circle*{0.1}}\put(27.12500,  5.388){\circle*{0.1}}
\put(27.25000,  5.955){\circle*{0.1}}\put(27.25000,  5.380){\circle*{0.1}}
\put(27.37500,  5.926){\circle*{0.1}}\put(27.37500,  5.371){\circle*{0.1}}
\put(27.50000,  5.898){\circle*{0.1}}\put(27.50000,  5.362){\circle*{0.1}}
\put(27.62500,  5.869){\circle*{0.1}}\put(27.62500,  5.352){\circle*{0.1}}
\put(27.75000,  5.840){\circle*{0.1}}\put(27.75000,  5.341){\circle*{0.1}}
\put(27.87500,  5.812){\circle*{0.1}}\put(27.87500,  5.329){\circle*{0.1}}
\put(28.00000,  5.783){\circle*{0.1}}\put(28.00000,  5.317){\circle*{0.1}}
\put(28.12500,  5.754){\circle*{0.1}}\put(28.12500,  5.304){\circle*{0.1}}
\put(28.25000,  5.725){\circle*{0.1}}\put(28.25000,  5.290){\circle*{0.1}}
\put(28.37500,  5.696){\circle*{0.1}}\put(28.37500,  5.275){\circle*{0.1}}
\put(28.50000,  5.667){\circle*{0.1}}\put(28.50000,  5.260){\circle*{0.1}}
\put(28.62500,  5.638){\circle*{0.1}}\put(28.62500,  5.245){\circle*{0.1}}
\put(28.75000,  5.609){\circle*{0.1}}\put(28.75000,  5.229){\circle*{0.1}}
\put(28.87500,  5.580){\circle*{0.1}}\put(28.87500,  5.212){\circle*{0.1}}
\put(29.00000,  5.550){\circle*{0.1}}\put(29.00000,  5.195){\circle*{0.1}}
\put(29.12500,  5.521){\circle*{0.1}}\put(29.12500,  5.177){\circle*{0.1}}
\put(29.25000,  5.491){\circle*{0.1}}\put(29.25000,  5.159){\circle*{0.1}}
\put(29.37500,  5.461){\circle*{0.1}}\put(29.37500,  5.140){\circle*{0.1}}
\put(29.50000,  5.431){\circle*{0.1}}\put(29.50000,  5.121){\circle*{0.1}}
\put(29.62500,  5.401){\circle*{0.1}}\put(29.62500,  5.101){\circle*{0.1}}
\put(29.75000,  5.370){\circle*{0.1}}\put(29.75000,  5.082){\circle*{0.1}}
\put(29.87500,  5.340){\circle*{0.1}}\put(29.87500,  5.061){\circle*{0.1}}
\put(30.00000,  5.309){\circle*{0.1}}\put(30.00000,  5.041){\circle*{0.1}}
\put(30.12500,  5.278){\circle*{0.1}}\put(30.12500,  5.020){\circle*{0.1}}
\put(30.25000,  5.247){\circle*{0.1}}\put(30.25000,  4.999){\circle*{0.1}}
\put(30.37500,  5.215){\circle*{0.1}}\put(30.37500,  4.977){\circle*{0.1}}
\put(30.50000,  5.184){\circle*{0.1}}\put(30.50000,  4.956){\circle*{0.1}}
\put(30.62500,  5.152){\circle*{0.1}}\put(30.62500,  4.934){\circle*{0.1}}
\put(30.75000,  5.121){\circle*{0.1}}\put(30.75000,  4.912){\circle*{0.1}}
\put(30.87500,  5.089){\circle*{0.1}}\put(30.87500,  4.889){\circle*{0.1}}
\put(31.00000,  5.057){\circle*{0.1}}\put(31.00000,  4.867){\circle*{0.1}}
\put(31.12500,  5.024){\circle*{0.1}}\put(31.12500,  4.844){\circle*{0.1}}
\put(31.25000,  4.992){\circle*{0.1}}\put(31.25000,  4.822){\circle*{0.1}}
\put(31.37500,  4.959){\circle*{0.1}}\put(31.37500,  4.799){\circle*{0.1}}
\put(31.50000,  4.927){\circle*{0.1}}\put(31.50000,  4.776){\circle*{0.1}}
\put(31.62500,  4.894){\circle*{0.1}}\put(31.62500,  4.753){\circle*{0.1}}
\put(31.75000,  4.861){\circle*{0.1}}\put(31.75000,  4.730){\circle*{0.1}}
\put(31.87500,  4.828){\circle*{0.1}}\put(31.87500,  4.706){\circle*{0.1}}
\put(32.00000,  4.795){\circle*{0.1}}\put(32.00000,  4.683){\circle*{0.1}}
\put(32.12500,  4.762){\circle*{0.1}}\put(32.12500,  4.660){\circle*{0.1}}
\put(32.25000,  4.728){\circle*{0.1}}\put(32.25000,  4.637){\circle*{0.1}}
\put(32.37500,  4.695){\circle*{0.1}}\put(32.37500,  4.613){\circle*{0.1}}
\put(32.50000,  4.661){\circle*{0.1}}\put(32.50000,  4.590){\circle*{0.1}}
\put(32.62500,  4.628){\circle*{0.1}}\put(32.62500,  4.567){\circle*{0.1}}
\put(32.75000,  4.594){\circle*{0.1}}\put(32.75000,  4.544){\circle*{0.1}}
\put(32.87500,  4.561){\circle*{0.1}}\put(32.87500,  4.521){\circle*{0.1}}
\put(33.00000,  4.527){\circle*{0.1}}\put(33.00000,  4.498){\circle*{0.1}}
\put(33.12500,  4.493){\circle*{0.1}}\put(33.12500,  4.475){\circle*{0.1}}
\put(33.25000,  4.459){\circle*{0.1}}\put(33.25000,  4.452){\circle*{0.1}}
\put(33.37500,  4.426){\circle*{0.1}}\put(33.37500,  4.429){\circle*{0.1}}
\put(33.50000,  4.392){\circle*{0.1}}\put(33.50000,  4.406){\circle*{0.1}}
\put(33.62500,  4.358){\circle*{0.1}}\put(33.62500,  4.384){\circle*{0.1}}
\put(33.75000,  4.324){\circle*{0.1}}\put(33.75000,  4.361){\circle*{0.1}}
\put(33.87500,  4.291){\circle*{0.1}}\put(33.87500,  4.339){\circle*{0.1}}
\put(34.00000,  4.257){\circle*{0.1}}\put(34.00000,  4.317){\circle*{0.1}}
\put(34.12500,  4.223){\circle*{0.1}}\put(34.12500,  4.295){\circle*{0.1}}
\put(34.25000,  4.190){\circle*{0.1}}\put(34.25000,  4.273){\circle*{0.1}}
\put(34.37500,  4.156){\circle*{0.1}}\put(34.37500,  4.251){\circle*{0.1}}
\put(34.50000,  4.123){\circle*{0.1}}\put(34.50000,  4.229){\circle*{0.1}}
\put(34.62500,  4.089){\circle*{0.1}}\put(34.62500,  4.207){\circle*{0.1}}
\put(34.75000,  4.056){\circle*{0.1}}\put(34.75000,  4.186){\circle*{0.1}}
\put(34.87500,  4.023){\circle*{0.1}}\put(34.87500,  4.165){\circle*{0.1}}
\put(35.00000,  3.990){\circle*{0.1}}\put(35.00000,  4.143){\circle*{0.1}}
\put(35.12500,  3.957){\circle*{0.1}}\put(35.12500,  4.122){\circle*{0.1}}
\put(35.25000,  3.924){\circle*{0.1}}\put(35.25000,  4.101){\circle*{0.1}}
\put(35.37500,  3.891){\circle*{0.1}}\put(35.37500,  4.081){\circle*{0.1}}
\put(35.50000,  3.858){\circle*{0.1}}\put(35.50000,  4.060){\circle*{0.1}}
\put(35.62500,  3.826){\circle*{0.1}}\put(35.62500,  4.040){\circle*{0.1}}
\put(35.75000,  3.794){\circle*{0.1}}\put(35.75000,  4.020){\circle*{0.1}}
\put(35.87500,  3.761){\circle*{0.1}}\put(35.87500,  3.999){\circle*{0.1}}
\put(36.00000,  3.729){\circle*{0.1}}\put(36.00000,  3.980){\circle*{0.1}}
\put(36.12500,  3.698){\circle*{0.1}}\put(36.12500,  3.960){\circle*{0.1}}
\put(36.25000,  3.666){\circle*{0.1}}\put(36.25000,  3.940){\circle*{0.1}}
\put(36.37500,  3.635){\circle*{0.1}}\put(36.37500,  3.921){\circle*{0.1}}
\put(36.50000,  3.603){\circle*{0.1}}\put(36.50000,  3.901){\circle*{0.1}}
\put(36.62500,  3.572){\circle*{0.1}}\put(36.62500,  3.882){\circle*{0.1}}
\put(36.75000,  3.541){\circle*{0.1}}\put(36.75000,  3.863){\circle*{0.1}}
\put(36.87500,  3.511){\circle*{0.1}}\put(36.87500,  3.844){\circle*{0.1}}
\put(37.00000,  3.480){\circle*{0.1}}\put(37.00000,  3.825){\circle*{0.1}}
\put(37.12500,  3.450){\circle*{0.1}}\put(37.12500,  3.807){\circle*{0.1}}
\put(37.25000,  3.420){\circle*{0.1}}\put(37.25000,  3.788){\circle*{0.1}}
\put(37.37500,  3.390){\circle*{0.1}}\put(37.37500,  3.770){\circle*{0.1}}
\put(37.50000,  3.360){\circle*{0.1}}\put(37.50000,  3.752){\circle*{0.1}}
\put(37.62500,  3.331){\circle*{0.1}}\put(37.62500,  3.734){\circle*{0.1}}
\put(37.75000,  3.302){\circle*{0.1}}\put(37.75000,  3.716){\circle*{0.1}}
\put(37.87500,  3.273){\circle*{0.1}}\put(37.87500,  3.698){\circle*{0.1}}
\put(38.00000,  3.244){\circle*{0.1}}\put(38.00000,  3.680){\circle*{0.1}}
\put(38.12500,  3.215){\circle*{0.1}}\put(38.12500,  3.663){\circle*{0.1}}
\put(38.25000,  3.187){\circle*{0.1}}\put(38.25000,  3.645){\circle*{0.1}}
\put(38.37500,  3.159){\circle*{0.1}}\put(38.37500,  3.628){\circle*{0.1}}
\put(38.50000,  3.131){\circle*{0.1}}\put(38.50000,  3.611){\circle*{0.1}}
\put(38.62500,  3.104){\circle*{0.1}}\put(38.62500,  3.593){\circle*{0.1}}
\put(38.75000,  3.076){\circle*{0.1}}\put(38.75000,  3.576){\circle*{0.1}}
\put(38.87500,  3.049){\circle*{0.1}}\put(38.87500,  3.559){\circle*{0.1}}
\put(39.00000,  3.022){\circle*{0.1}}\put(39.00000,  3.543){\circle*{0.1}}
\put(39.12500,  2.996){\circle*{0.1}}\put(39.12500,  3.526){\circle*{0.1}}
\put(39.25000,  2.969){\circle*{0.1}}\put(39.25000,  3.509){\circle*{0.1}}
\put(39.37500,  2.943){\circle*{0.1}}\put(39.37500,  3.493){\circle*{0.1}}
\put(39.50000,  2.917){\circle*{0.1}}\put(39.50000,  3.476){\circle*{0.1}}
\put(39.62500,  2.892){\circle*{0.1}}\put(39.62500,  3.460){\circle*{0.1}}
\put(39.75000,  2.866){\circle*{0.1}}\put(39.75000,  3.443){\circle*{0.1}}
\put(39.87500,  2.841){\circle*{0.1}}\put(39.87500,  3.427){\circle*{0.1}}
\put(40.00000,  2.816){\circle*{0.1}}\put(40.00000,  3.411){\circle*{0.1}}
\put(40.12500,  2.791){\circle*{0.1}}\put(40.12500,  3.395){\circle*{0.1}}
\put(40.25000,  2.767){\circle*{0.1}}\put(40.25000,  3.379){\circle*{0.1}}
\put(40.37500,  2.743){\circle*{0.1}}\put(40.37500,  3.363){\circle*{0.1}}
\put(40.50000,  2.719){\circle*{0.1}}\put(40.50000,  3.347){\circle*{0.1}}
\put(40.62500,  2.695){\circle*{0.1}}\put(40.62500,  3.331){\circle*{0.1}}
\put(40.75000,  2.671){\circle*{0.1}}\put(40.75000,  3.315){\circle*{0.1}}
\put(40.87500,  2.648){\circle*{0.1}}\put(40.87500,  3.300){\circle*{0.1}}
\put(41.00000,  2.625){\circle*{0.1}}\put(41.00000,  3.284){\circle*{0.1}}
\put(41.12500,  2.602){\circle*{0.1}}\put(41.12500,  3.268){\circle*{0.1}}
\put(41.25000,  2.579){\circle*{0.1}}\put(41.25000,  3.253){\circle*{0.1}}
\put(41.37500,  2.557){\circle*{0.1}}\put(41.37500,  3.237){\circle*{0.1}}
\put(41.50000,  2.535){\circle*{0.1}}\put(41.50000,  3.222){\circle*{0.1}}
\put(41.62500,  2.513){\circle*{0.1}}\put(41.62500,  3.206){\circle*{0.1}}
\put(41.75000,  2.491){\circle*{0.1}}\put(41.75000,  3.191){\circle*{0.1}}
\put(41.87500,  2.470){\circle*{0.1}}\put(41.87500,  3.176){\circle*{0.1}}
\put(42.00000,  2.448){\circle*{0.1}}\put(42.00000,  3.160){\circle*{0.1}}
\put(42.12500,  2.427){\circle*{0.1}}\put(42.12500,  3.145){\circle*{0.1}}
\put(42.25000,  2.406){\circle*{0.1}}\put(42.25000,  3.130){\circle*{0.1}}
\put(42.37500,  2.385){\circle*{0.1}}\put(42.37500,  3.115){\circle*{0.1}}
\put(42.50000,  2.365){\circle*{0.1}}\put(42.50000,  3.100){\circle*{0.1}}
\put(42.62500,  2.345){\circle*{0.1}}\put(42.62500,  3.084){\circle*{0.1}}
\put(42.75000,  2.325){\circle*{0.1}}\put(42.75000,  3.069){\circle*{0.1}}
\put(42.87500,  2.305){\circle*{0.1}}\put(42.87500,  3.054){\circle*{0.1}}
\put(43.00000,  2.285){\circle*{0.1}}\put(43.00000,  3.039){\circle*{0.1}}
\put(43.12500,  2.265){\circle*{0.1}}\put(43.12500,  3.024){\circle*{0.1}}
\put(43.25000,  2.246){\circle*{0.1}}\put(43.25000,  3.010){\circle*{0.1}}
\put(43.37500,  2.227){\circle*{0.1}}\put(43.37500,  2.995){\circle*{0.1}}
\put(43.50000,  2.208){\circle*{0.1}}\put(43.50000,  2.980){\circle*{0.1}}
\put(43.62500,  2.189){\circle*{0.1}}\put(43.62500,  2.965){\circle*{0.1}}
\put(43.75000,  2.170){\circle*{0.1}}\put(43.75000,  2.950){\circle*{0.1}}
\put(43.87500,  2.152){\circle*{0.1}}\put(43.87500,  2.935){\circle*{0.1}}
\put(44.00000,  2.134){\circle*{0.1}}\put(44.00000,  2.921){\circle*{0.1}}
\put(44.12500,  2.116){\circle*{0.1}}\put(44.12500,  2.906){\circle*{0.1}}
\put(44.25000,  2.098){\circle*{0.1}}\put(44.25000,  2.891){\circle*{0.1}}
\put(44.37500,  2.080){\circle*{0.1}}\put(44.37500,  2.877){\circle*{0.1}}
\put(44.50000,  2.062){\circle*{0.1}}\put(44.50000,  2.862){\circle*{0.1}}
\put(44.62500,  2.045){\circle*{0.1}}\put(44.62500,  2.847){\circle*{0.1}}
\put(44.75000,  2.027){\circle*{0.1}}\put(44.75000,  2.833){\circle*{0.1}}
\put(44.87500,  2.010){\circle*{0.1}}\put(44.87500,  2.818){\circle*{0.1}}
\put(45.00000,  1.993){\circle*{0.1}}\put(45.00000,  2.804){\circle*{0.1}}
\put(45.12500,  1.976){\circle*{0.1}}\put(45.12500,  2.789){\circle*{0.1}}
\put(45.25000,  1.960){\circle*{0.1}}\put(45.25000,  2.775){\circle*{0.1}}
\put(45.37500,  1.943){\circle*{0.1}}\put(45.37500,  2.760){\circle*{0.1}}
\put(45.50000,  1.927){\circle*{0.1}}\put(45.50000,  2.746){\circle*{0.1}}
\put(45.62500,  1.910){\circle*{0.1}}\put(45.62500,  2.732){\circle*{0.1}}
\put(45.75000,  1.894){\circle*{0.1}}\put(45.75000,  2.717){\circle*{0.1}}
\put(45.87500,  1.878){\circle*{0.1}}\put(45.87500,  2.703){\circle*{0.1}}
\put(46.00000,  1.863){\circle*{0.1}}\put(46.00000,  2.689){\circle*{0.1}}
\put(46.12500,  1.847){\circle*{0.1}}\put(46.12500,  2.675){\circle*{0.1}}
\put(46.25000,  1.831){\circle*{0.1}}\put(46.25000,  2.661){\circle*{0.1}}
\put(46.37500,  1.816){\circle*{0.1}}\put(46.37500,  2.646){\circle*{0.1}}
\put(46.50000,  1.800){\circle*{0.1}}\put(46.50000,  2.632){\circle*{0.1}}
\put(46.62500,  1.785){\circle*{0.1}}\put(46.62500,  2.618){\circle*{0.1}}
\put(46.75000,  1.770){\circle*{0.1}}\put(46.75000,  2.604){\circle*{0.1}}
\put(46.87500,  1.755){\circle*{0.1}}\put(46.87500,  2.590){\circle*{0.1}}
\put(47.00000,  1.741){\circle*{0.1}}\put(47.00000,  2.576){\circle*{0.1}}
\put(47.12500,  1.726){\circle*{0.1}}\put(47.12500,  2.563){\circle*{0.1}}
\put(47.25000,  1.711){\circle*{0.1}}\put(47.25000,  2.549){\circle*{0.1}}
\put(47.37500,  1.697){\circle*{0.1}}\put(47.37500,  2.535){\circle*{0.1}}
\put(47.50000,  1.682){\circle*{0.1}}\put(47.50000,  2.521){\circle*{0.1}}
\put(47.62500,  1.668){\circle*{0.1}}\put(47.62500,  2.507){\circle*{0.1}}
\put(47.75000,  1.654){\circle*{0.1}}\put(47.75000,  2.494){\circle*{0.1}}
\put(47.87500,  1.640){\circle*{0.1}}\put(47.87500,  2.480){\circle*{0.1}}
\put(48.00000,  1.626){\circle*{0.1}}\put(48.00000,  2.467){\circle*{0.1}}
\put(48.12500,  1.613){\circle*{0.1}}\put(48.12500,  2.453){\circle*{0.1}}
\put(48.25000,  1.599){\circle*{0.1}}\put(48.25000,  2.440){\circle*{0.1}}
\put(48.37500,  1.585){\circle*{0.1}}\put(48.37500,  2.426){\circle*{0.1}}
\put(48.50000,  1.572){\circle*{0.1}}\put(48.50000,  2.413){\circle*{0.1}}
\put(48.62500,  1.559){\circle*{0.1}}\put(48.62500,  2.399){\circle*{0.1}}
\put(48.75000,  1.545){\circle*{0.1}}\put(48.75000,  2.386){\circle*{0.1}}
\put(48.87500,  1.532){\circle*{0.1}}\put(48.87500,  2.373){\circle*{0.1}}
\put(49.00000,  1.519){\circle*{0.1}}\put(49.00000,  2.360){\circle*{0.1}}
\put(49.12500,  1.506){\circle*{0.1}}\put(49.12500,  2.347){\circle*{0.1}}
\put(49.25000,  1.494){\circle*{0.1}}\put(49.25000,  2.333){\circle*{0.1}}
\put(49.37500,  1.481){\circle*{0.1}}\put(49.37500,  2.320){\circle*{0.1}}
\put(49.50000,  1.468){\circle*{0.1}}\put(49.50000,  2.307){\circle*{0.1}}
\put(49.62500,  1.456){\circle*{0.1}}\put(49.62500,  2.294){\circle*{0.1}}
\put(49.75000,  1.443){\circle*{0.1}}\put(49.75000,  2.282){\circle*{0.1}}
\put(49.87500,  1.431){\circle*{0.1}}\put(49.87500,  2.269){\circle*{0.1}}
\put(50.00000,  1.419){\circle*{0.1}}\put(50.00000,  2.256){\circle*{0.1}}
\end{picture}%
}
\put(145,25){%
\begin{picture}(0,0)%
\put(-40,0){\line(1,0){70}}
\put(0,0){\line(0,1){70}}
\multiput(-40,-2)(10,0){8}{\line(0,1){2}}
\multiput(-2,10)(0,10){7}{\line(1,0){2}}
\put(40,-10){\llap{${\rm Re}\,a_{\eta A}$}}
\put(-2,75){\llap{${\rm Im}\,a_{\eta A}$}}
\put(-44,-10){-4}
\put(-34,-10){-3}
\put(-24,-10){-2}
\put(-14,-10){-1}
\put(-2,-10){0}
\put(8,-10){1}
\put(17,12){${}^2{\rm H}$}
\put(-42,35){${}^4{\rm He}$}
\put(5.60,11.91){\circle{3.00}}
\put(8.73,15.07){\circle{3.00}}
\put(12.10,19.66){\circle{3.00}}
\put(15.22,26.43){\circle{3.00}}
\put(16.64,36.21){\circle{3.00}}
\put(13.40,48.53){\circle{3.00}}
\put(3.01,59.18){\circle{3.00}}
\put(-11.25,62.50){\circle*{3.00}}
\put(-11.91,31.00){\circle{3.00}}
\put(-26.13,32.69){\circle{3.00}}
\put(-35.73,23.09){\circle*{3.00}}
\put(-35.94,13.60){\circle*{3.00}}
\put(-32.98,8.45){\circle*{3.00}}
\put(-30.05,6.33){\circle*{3.00}}
\put(-28.42,5.50){\circle*{3.00}}
\put(-26.41,4.27){\circle*{3.00}}
\put(-10,-25){Fig. 2}
\end{picture}%
}
\put(255,25){%
\begin{picture}(0,0)%
\put(-40,0){\line(1,0){70}}
\put(0,0){\line(0,1){70}}
\multiput(-40,-2)(10,0){8}{\line(0,1){2}}
\multiput(-2,10)(0,10){7}{\line(1,0){2}}
\put(40,-10){\llap{${\rm Re}\,a_{\eta A}$}}
\put(-2,75){\llap{${\rm Im}\,a_{\eta A}$}}
\put(-44,-10){-4}
\put(-34,-10){-3}
\put(-24,-10){-2}
\put(-14,-10){-1}
\put(-2,-10){0}
\put(8,-10){1}
\put(-35,47){${}^3{\rm H}$}
\put(-25,33){${}^3{\rm He}$}
\put(2.21,23.48){\circle{3.00}}
\put(0.82,33.01){\circle{3.00}}
\put(-7.38,43.13){\circle{3.00}}
\put(-22.07,46.1){\circle{3.00}}
\put(-33.65,39.41){\circle{3.00}}
\put(-37.84,30.71){\circle*{3.00}}
\put(-37.99,24.28){\circle*{3.00}}
\put(-36.83,20.19){\circle*{3.00}}
\put(2.03,23.43){\circle{3.00}}
\put(0.60,32.80){\circle{3.00}}
\put(-7.42,42.63){\circle{3.00}}
\put(-21.57,45.60){\circle{3.00}}
\put(-37.26,31.17){\circle*{3.00}}
\put(-37.69,24.92){\circle*{3.00}}
\put(-36.77,20.88){\circle*{3.00}}
\put(-10,-25){Fig. 3}
\end{picture}%
}
\end{picture}
\end{center}

The FRA approach enables us to look for bound and resonant states as
singularities of the $S$--matrix. It is known \cite{Cass} that the
resonant and quasi--bound state poles of the $S$--matrix, generated
by a non-Hermitian Hamiltonian, are situated in the second quadrant
of the complex momentum plane (below and above its diagonal
respectively). The poles found with $a_{\eta N}=(0.55+i0.30)$ fm are
shown in Fig. 4. When ${\rm Re\,}a_{\eta N}$ increases, all the poles
move up and to the right, and when a resonance pole crosses the
diagonal it becomes a quasi-bound pole. The minimal values of
${\rm Re\,}a_{\eta N}$ which generate the `zero--binding'
(the poles just on the diagonal) are given in the table below.\\

\unitlength=0.5mm
\begin{picture}(280,90)
\put(100,35){%
\begin{picture}(0,0)%
\put(0,0){\line(-1,0){100}}
\put(0,0){\line(0,1){40}}
\multiput(0,-2)(-10,0){11}{\line(0,1){2}}
\put(0,-4){\line(0,1){2}}
\put(-50,-4){\line(0,1){2}}
\put(-100,-4){\line(0,1){2}}
\multiput(2,0)(0,10){5}{\line(-1,0){2}}
\put(-20,45){${\rm Im\,}p\, ({\rm fm}^{-1})$}
\put(-105,-15){${\rm Re\,}p\, ({\rm fm}^{-1})$}
\put(-58,-12){-0.5}
\put(0,0){\line(-1,1){40}}
\put(4,8){0.1}
\put(4,18){0.2}
\put(4,28){0.3}
\put(4,-2){0}
\put(-1,-12){0}
\put(-50,-30){Fig. 4}
\put(-90.259,35.870){\circle{3.00}}
\put(-56.045,23.859){\circle{3.00}}
\put(-54.692,24.478){\circle{3.00}}
\put(-16.504,27.876){\circle*{3.00}}
\put(-90,40){${}^2{\rm H}$}
\put(-70,20){${}^3{\rm H}$}
\put(-50,25){${}^3{\rm He}$}
\put(-15,17){${}^4{\rm He}$}
\put(33,15){%
\begin{tabular}{|c|c|c|c|c|}
\hline
system & $\phantom{-}\eta {}^2{\rm H}^{\mathstrut}_{\mathstrut}$
\phantom{-}& $\phantom{-}\eta {}^3{\rm H}$\phantom{-} &
	 $\phantom{-}\eta {}^3{\rm He}$\phantom{-} &
	 $\phantom{-}\eta {}^4{\rm He}$\phantom{-} \\
\hline
min$\{{\rm Re\,}a_{\eta N}\}^{\mathstrut}_{\mathstrut}$\ (fm) &
\phantom{-}0.91\phantom{-} & \phantom{-}0.75\phantom{-} &
\phantom{-}0.73\phantom{-} &\phantom{-} 0.47\phantom{-}\\
\hline
\end{tabular}%
}
\end{picture}%
}
\put(220,5){Table 1}
\end{picture}

All these values are within the uncertainty interval
${\rm Re\,}a_{\eta N}\in [0.27,0.98]$ fm. Thus even the possibility
of an $\eta$d binding cannot be at present excluded. Latest
estimates of ${\rm Re\,}a_{\eta N}$ \cite{newa} are concentrated
around the value ${\rm Re\,}a_{\eta N}\approx 0.7$ fm, which enhances
our belief that at least the  $\alpha$--particle can entrap  an
$\eta$--meson.

Finally, we would like to emphasize that the  spectral properties of
Hermitian and non-Hermitian Hamiltonians are quite different. Locating
quasi--bound states is a delicate problem which can be treated only
by rigorous methods. As we have shown in Ref. \cite{Raki4} the
$\eta$A scattering length can say nothing about the existence or not
of an $\eta A$ quasi--bound state. This is clearly seen on Figs. 2
and 3 where the trajectories go smoothly from open to filled circles
without any drastic changes or extreme values.


\end{document}